\documentclass[pra,showpacs,showkeys]{revtex4}
\usepackage{graphicx}
\usepackage{indentfirst}
\usepackage{dcolumn}
\usepackage{amsfonts}
\usepackage{amsmath}
\usepackage{bm}
\newcommand{\ket}[1]{\left|#1\right\rangle}
\newcommand{\bra}[1]{\left\langle#1\right|}
\newcommand{\bgeq}{\begin{equation}}
\newcommand{\edeq}{\end{equation}}
\newcommand{\bgeqn}{\begin{eqnarray}}
\newcommand{\edeqn}{\end{eqnarray}}
\newcommand{\tr}[1]{{\rm Tr\{ #1 \}}}
\newtheorem{theorem}{Proposition}

\newtheorem{corollary}[theorem]{Corollary}

\begin{document}

\title{Correlations in local measurements on a quantum state, and
complementarity as an explanation of nonclassicality }

\author{Shengjun Wu$^{1,2}$}
\author{Uffe V. Poulsen$^1$}
\author{Klaus M{\o}lmer$^1$}
\affiliation{
$^1$Lundbeck Foundation Theoretical Center for Quantum System Research,
Department of Physics and Astronomy, Aarhus University, DK-8000 Aarhus C, Denmark
\\
$^2$Hefei National Laboratory for Physical Sciences at Microscale, \\
University of Science and Technology of China, Hefei, Anhui 230026, P. R. China
}

\date{\today}

\begin{abstract}
We consider the classical correlations that two observers can extract by measurements on
a bipartite quantum state, and we discuss how they are related to the quantum mutual information of
the state. We show with several examples how complementarity
gives rise to a gap between the quantum and the classical correlations, and we relate our
quantitative finding to the so-called classical correlation locked in a quantum state.
We derive upper bounds for the sum of classical correlation
obtained by measurements in different mutually unbiased bases and we show that
the complementarity gap is also present in the
deterministic quantum computation with one quantum bit.
\end{abstract}

\pacs{03.65.Ud, 03.67.Hk, 42.50.Dv, 03.67.Ac}
\keywords{nonclassical correlation, quantum measurements, complementarity, quantum communication, quantum computing}
\maketitle

\section{Introduction}

Complementarity was identified by Niels Bohr as the fundamental
underlying cause of quantum uncertainty and as an unavoidable element
in our attempts to simultaneously describe different properties of
physical systems \cite{Bohr}. Formally, complementarity is reflected by the
non-commuting mathematical operators representing physical observables
which prevent the general existence of joint eigenstates of these
operators, and which hence leads to uncertain measurement
outcomes. Complementarity is also reflected in the impossibility for
any physical set-up to measure such different observables without the
measurement of one observable disturbing the outcome of the
other. In an attempt to challenge the
complementarity view, Einstein, Podolsky and Rosen, introduced the
entangled EPR-state \cite{EPR} in which quantum correlations between the
constituents are strong enough to suggest an existence of
action-at-a-distance, when measurements on one particle seemingly
force the projection of the quantum state of the other particle. This
effect has been a corner stone in the discussion of the interpretation
of quantum mechanics, but rather than weakening Bohr's views it served
to strengthen the complementarity description, and more recently it
has been incorporated as a useful paradigm and even a practical
ingredient in quantum information processing tasks. The EPR state can thus be used for sharing of a secret key between remote partners \cite{crypto} and for teleportation of unknown quantum states \cite{teleportation}, and it can assist in various other communication \cite{Wiesner,frames} and precision measurement tasks \cite{precision}.

With the numerous applications of such correlated states, it has been
a natural goal to quantify the degree of correlation of any given
state, and also to quantify to which extent the correlations are of a
quantum or of a classical nature. Entanglement does not exist in
classical physics, and part of this classification has therefore been
related to the ability to distinguish entangled states from
non-entangled states and, more ambitiously, to quantify the amount of
entanglement in a given quantum state. This is a research field where
considerable progress has been made, but where many open and difficult
questions still remain to be solved.

Another natural approach to the characterization of the quantum
correlations in a given state can be based upon the comparison of the
information theoretic correlations of the quantum state with the
classical correlations present in the classical detection records
after measurements have taken
place~\cite{Luo082,OZ01,CA97,PhysRevLett.91.117901,PhysRevA.71.062307,PhysRevA.72.032317,horodecki05:_partial_quant_info,devetak04:_distil_common_random_quant_states}. Studies
have revealed that some quantum states carry more correlations
than can be retrieved by separate measurements on the constituents,
even if they are not entangled; they may posses powers for quantum
information tasks which are not available in classical states, and by
allowing a parallel classical communication one may extract or
transfer more information with such states than the sum of the
classical communication and the previously available information. A
number of natural quantitative measures of quantumness have thus
appeared, which in one way or another display the difference between
the correlations present in the original quantum state, and the
information retrieved by suitable schemes or protocols. In this paper
we shall suggest that this difference may be ascribed to
complementarity, and we quantify how
different measurement strategies provide upper and lower bounds for
different information characteristics.

In Sec. II, we introduce the quantum mutual information of a quantum
state and the maximal mutual information between classical measurement
records obtained by measurements on the quantum state. The difference
between these two quantities constitutes a measure $Q$ of the
quantumness of the correlations of the quantum state, and we shall
present some quantitative results obeyed by $Q$. In Sec. III, we
discuss the role of complementarity when classical information is
extracted from quantum states, and we relate our complementarity
discussion and $Q$ to the so-called locking effect. In Sec. IV, we
review the properties of two other measures of non-classical
correlations: the quantum discord and the measurement induced
disturbance, and we establish the relationship between them and our
$Q$. In Sec. V we present a number of examples which quantitatively illustrate our results. In Sec. VI, we discuss consequences of the complementarity on
classical correlations obtained with projective measurements on
mutually unbiased bases. In Sec. VII, we show how our measure of
quantumness quantifies the speed-up of quantum computing in a
particular model where there is no distillable entanglement
present. Sec. VIII concludes the paper.

\section{Quantum mutual information and
classical correlations  generated by local measurements
on a quantum state}

\begin{figure}[tbp]
  \centering
  \resizebox{0.9\textwidth}{!}{
    \includegraphics{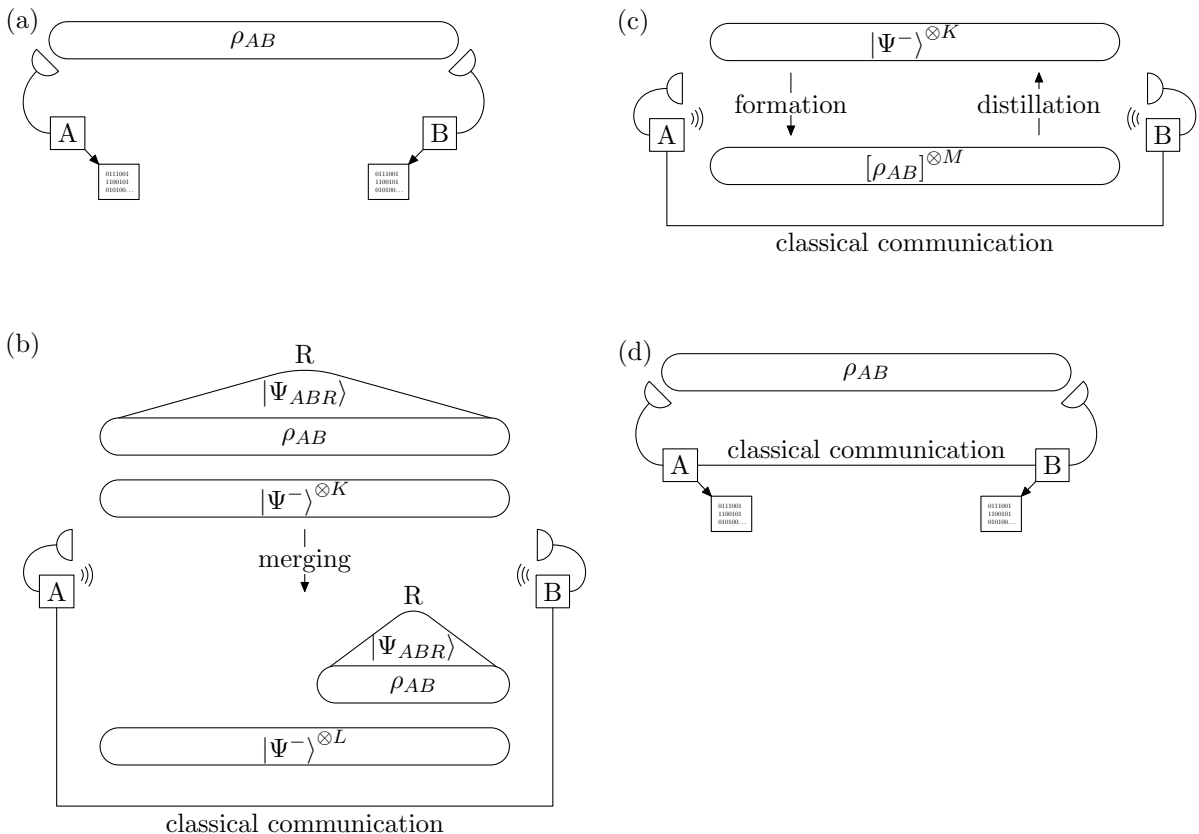}
  }
  \caption{Four quantum information processing protocols involving shared bipartite quantum states: (a)
    Extracting classical measurement records by local measurements
    with the aim to maximize the mutual information between the
    records. (b) Merging Alice's
    part of the state to Bob while preserving any correlations with a
    (possibly fictitious) reservoir R. The merging uses local operations and classical communication, and it may consume $(K>L)$ or, actually, produce $(L>K)$ maximally entangled states between Alice and Bob.(c) Interconverting between $M$ copies of a
    quantum state and $K$ maximally entangled qubit pairs, $|\Psi^-\rangle=\frac{1}{\sqrt{2}}(|0_A1_B\rangle - |1_A0_B\rangle$,
    while allowing classical communication. (d) Extracting classical measurement records while
    allowing classical communication. This may increase the mutual
    information by a larger amount than what is transmitted  through the
    classical channel, an effect referred to as ``locking'' of
    classical information in a quantum state (see Sec. IIIB).}
  \label{fig:protocols}
\end{figure}

Alice and Bob share a quantum state described by the density matrix
$\rho_{AB}$, with the marginal states denoted by $\rho_A\equiv
\textrm{Tr}_B(\rho_{AB})$ and $\rho_B\equiv \textrm{Tr}_A(\rho_{AB})$.
The most general operations performed by Alice and Bob \emph{without}
any communication between them are completely positive, local
maps, in which Alice performs a general POVM measurement ($\{M_i^A |
M_i^A\geq 0,\sum_i M_i^A=I_A\}$) on her part of the quantum system and
Bob performs a general measurement ($\{M_s^B| M_s^B \geq 0,\sum_s
M_s^B=I_B\}$) on his part of the system,
see~Fig.\ref{fig:protocols}(a). The marginal probability
$p_i=\tr{(M_i^A\otimes I_B) \rho_{AB}}=\tr{M_i^A \rho_A}$
($p_s=\tr{(I_A\otimes M_s^B) \rho_{AB}}=\tr{M_s^B \rho_B}$) denotes
the probability of obtaining the $i$th ($s$th) result by Alice (Bob)
regardless of the other observer's result, while the joint probability
that Alice obtains the $i$th result and Bob obtains the $s$th result
is given by $p_{is}=\tr{(M_i^A\otimes M_s^B) \rho_{AB}}$. The
correlations in the two classical detection records obtained by Alice
and Bob are characterized by the mutual information $I\{A:B\}$,
defined by \bgeq I\{A:B\} = H\{ p_i; i \} + H\{ p_s; s\} - H\{ p_{is};
is \} \edeq where $H\{ p_\mu; \mu \}\equiv - \sum_\mu p_\mu \log_2
p_\mu$ denotes the Shannon entropies for the marginal probability
distribution $\{p_\mu= p_i,p_s\}$ and the joint probability
distribution $\{p_\mu= p_{is}\}$, respectively.

Without recourse to any particular measurement protocol, one defines
the quantum mutual information $S(A:B)$ of the state $\rho_{AB}$ in a
very similar manner, but in terms of the von Neumann entropies
$S(\rho)=-\textrm{Tr}[\rho \textrm{log}_2(\rho)]$ of $\rho_{AB}$,
$\rho_A$ and $\rho_B$,
\begin{equation}
S(A:B)=S(\rho_A)+S(\rho_B)-S(\rho_{AB}).
\end{equation}

The quantum mutual information $S(A:B)$ is readily evaluated for any
quantum state $\rho_{AB}$, and an important contribution to its precise operational meaning is provided in
Ref.~\cite{PhysRevA.72.032317} where it is shown that $S(A:B)$ quantifies the minimal amount of noise to be added
to a state to bring it to product form, i.e.\ to delete \emph{all}
quantum and classical correlations present in the state. Also, recent work on the so-called
merging effect has identified the difference $S(A:B)-S(A)$
as the number of maximally entangled qubit pairs that must be spent or
are released upon merging of the joint quantum state of $A$ and $B$ at
Bob, retaining their joint correlations with any ancilla degrees of
freedom~\cite{horodecki05:_partial_quant_info}, see Fig.~\ref{fig:protocols}(b).

For a pure quantum state, the density matrix can be replaced by a
state vector $|\psi_{AB}\rangle$, and $S(\rho_A)=S(\rho_B)$ quantifies
the entanglement of the state. $S(\rho_A)$ is for example the
asymptotic ratio $K/M$ between the number (K) of maximally entangled
qubit states, $|\Psi^-\rangle = \frac{1}{\sqrt{2}}(|0_A1_B\rangle -
|1_A0_B\rangle)$, that can be exchanged for $M$ copies of the state
$|\psi_{AB}\rangle$ using only local operations and classic
communication (LOCC), see Fig.~\ref{fig:protocols}(c)
\cite{BBPS96}. $S(AB)$ vanishes for a pure state, and hence
$S(A:B)=2S(\rho_A)$ also becomes a quantitative measure of the
entanglement of the state.
For a generally mixed state $\rho_{AB}$, the quantum mutual information of a $n$-copy broadcast state of $\rho_{AB}$,
i.e., a state of $n$ copies of systems $A_i$ and $B_i$ ($i=1,\cdots,n$) so that the reduced density matrix for
each pair ($A_i$, $B_i$) is $\rho_{AB}$,
can also be used to classify classical, separable and entangled states. It is, e.g., shown in \cite{PCMH09}
that the minimal per-copy quantum mutual information for an infinite number of broadcast
copies satisfies some properties required of an entanglement measure.

Both the merging analysis and the quantitative entanglement measure in
terms of an equivalent number of maximally entangled states assume
that local operations and classical communication between Alice and
Bob are free of charge, i.e., they provide measures of some
correlations which are of a strictly non-classical nature, but they do
not properly account for the classical correlations already present in
the state. By comparing $S(A:B)$, and $S(\rho_A)$,
$S(\rho_B)$ with $I\{A:B\}$ we shall shed more light on the meaning of the
quantum mutual information, and we shall, in particular investigate to
what extent it limits the classical mutual information available under
different measurement protocols.

\begin{theorem} \label{the1}
The classical mutual information has the following upper bound in terms of the von Neumann entropy and the quantum mutual
information:
\bgeq
I\{A:B\} \leq min\{ S(\rho_A), S(\rho_B), S(A:B)\}.
\edeq
\end{theorem}

The Holevo bound \cite{Hol73, FC94, YO93} can be used to prove that
$I\{A:B\} \leq S(\rho_A)$ and $I\{A:B\} \leq S(\rho_B)$. These
inequalities are also intuitively reasonable as $S(\rho_A)$ and
$S(\rho_B)$ by the Schumacher noiseless channel coding theorem
\cite{Schumacher05} denotes the effective size of the system on
Alice and Bob's side respectively. A proof of the inequality
$I\{A:B\} \leq S(A:B)$ can be found in \cite{BP91}. An alternative
complete proof of the proposition is given in the Appendix of this
paper.

We will denote by $I_{max}(\rho_{AB})$ the maximal mutual information over all choices
of local measurement strategies, as it is a property of the state $\rho_{AB}$
\cite{DHLST04,Hall06}.
$I_{max}(\rho_{AB})$ can always be achieved by
local measurements with rank-one POVM elements,
since a general POVM measurement can be refined to a rank-one POVM measurement
via eigen-decomposition of each POVM element. The original measurement can thus be
considered as a two-step procedure where first, the refined rank-1 measurement is performed, and then the output is
binned according to the original POVM elements. The mutual information $I\{A:B\}$ cannot
increase during the second step, and therefore it is sufficient to maximize over only the local measurements with rank-one
POVM elements when we calculate $I_{max}(\rho_{AB})$.
$I_{max}(\rho_{AB})$ may not always be achievable by local projective measurements,
i.e., local measurements with \emph{orthonormal} rank-one POVM elements(see example E in Sec. V).

We will introduce the difference between $S(A:B)$ and
$I_{max}(\rho_{AB})$,
\bgeq
Q(A:B) = S(A:B) - I_{max}(\rho_{AB}) .
\edeq
$Q(A:B)$ is a function of the quantum state, and we will also
use the notation $Q(\rho_{AB})$. This difference was also introduced with the symbol
$\triangle_{CC}(\rho_{AB})$, and its vanishing for states that are locally broadcastable
was shown in \cite{PHH08}.

When $\rho_{AB}=\ket{\psi_{AB}} \bra{\psi_{AB}}$ is a pure state,
$S(\rho_{AB})=0$ and $S(\rho_A)=S(\rho_B)$ is the entanglement of the
state. Proposition \ref{the1} then implies that the classical mutual
information is limited by the entanglement of the state
$S(\rho_A)=S(\rho_B) < S(A:B)$.  When the Schmidt bases are chosen as
the measuring bases for projective measurements by both observers
$I\{A:B\}$ actually reaches this upper bound, and $Q(A:B)$ is equal to
the entanglement of the state. For a mixed state, the relation between
entanglement and $I_{max}$ is more complicated, as on the one hand the
entanglement measure itself is not unique, and as a mixed state may
contain also classical correlations.  Since $I_{max}$ can be nonzero
even for a separable state, it can be larger than the amount of
entanglement, but according to proposition \ref{the1}, it cannot be
larger than the quantum mutual information $S(A:B)$, which accounts
for both the quantum and the classical correlations in the state.

A main objective of this paper is to address to which extent a given
measurement strategy reaches the limits of Proposition 1, and in
particular to identify the properties of $\rho_{AB}$, that may prevent
$I_{max}(\rho_{AB})$ from actually reaching these limits. In
all our examples, we will show that complementarity arguments predict
whether a measurement strategy allows $I\{A:B\}$ to exhaust
Proposition 1 or not.

\section{Complementarity, locking effect}

In the previous section we defined measures of quantum correlations in quantum states and we compared them with the correlations in classical detection records. In this section we will argue that complementarity issues directly limit the available classical information. And we will show how the so-called locking effect can be interpreted in these terms.

\subsection{Complementarity gaps}

Although we will be primarily interested in the two-party situation where Alice and Bob have access to a joint quantum state, let us first consider the situation where one person wants to transmit a message by preparing and sending a quantum system to a receiver, who may perform measurements on the system. Quantum states are specified by amplitudes with arbitrary complex values, but it is well known that for example no more than a single classical bit can be transmitted by a two-level quantum system in such a protocol. The random outcome of any binary measurement on a qubit in fact gives even less information unless the sender and receiver agree on which orthogonal basis is used for the preparation and detection, and the optimal communication is therefore effectively classical.

It is interesting to see how the use of a higher dimensional quantum system and a biased alphabet,
where the sender chooses states from an ensemble with average density matrix $\rho$,
described by the von Neumann entropy $S(\rho)=-\textrm{Tr}[\rho \textrm{log}_2(\rho)]$,
can lead to transmission of maximally $S(\rho)$ classical bits per particle.
This works if the sender uses pure states from the eigenbasis of $\rho$ with the corresponding probabilities,
and if the receiver measures in the same basis. In that case the Shannon entropy
$H\{ p_i; i \}\equiv - \sum_i p_i \log_2 p_i$ of
the probabilities for the different outcomes equals precisely the von Neumann entropy.
If we associate to the von Neumann entropy the amount of information in the quantum state,
and to the Shannon entropy, the amount of information available in the detection records,
we see that these two measures agree, if the same observable is used for encoding and readout of the transmitted  message,
while if the observable used by the receiver is complementary to the one used by the sender,
the classical detection record does not exhaust the information available in the quantum state.

In the situation we are interested in, Alice and Bob share a quantum
system. Any observable addressed by Alice commutes with any observable
addressed by Bob, and hence complementarity does not prevent Alice and
Bob from measuring any observable of their choice with any desired
precision. Given the precise quantum state, however, there may be
correlations referring to specific observables, for example strong
correlations of the particle positions in the original EPR state or of
their spin components in the later spin version due to
Bohm. Experiments on the EPR states are known to violate Bell's
inequality (and a number of related inequalities), revealing that
measurements on the states cannot be described by local hidden
variable theories. We shall here argue that complementarity arguments
here also lead to inequalities that must be \textit{fulfilled} by
measurements on bipartite quantum states. To be a little more precise:
Let Alice perform, e.g., a projective measurement of an
observable $A$, and let the quantum state carry a correlation between
this observable and an observable $B$, enabling Alice to infer,
precisely or approximately, the outcome of a measurement of an
associated observable $B$ by Bob. Bob may, however, be interested in another observable $C$, which is
complementary to $B$, and hence
joint information about the formally commuting observables $A$ and $C$, \textit{de facto}, becomes
complementary. We will now proceed to show how the possible
complementarity between different observables that one would like to measure on
a subsystem, or between a single observable, actually measured on a
subsystem and observables inferred from measurements on the other
subsystem, is decisive for the existence of a gap between
the correlations present in a quantum state and the classical correlations
that can be extracted by local measurements.

When $\rho_{AB}=\ket{\psi_{AB}} \bra{\psi_{AB}}$ is a pure, entangled
state, $S(\rho_{AB})=0$ and $S(\rho_A)=S(\rho_B)$ is the entanglement
of the state. According to Proposition 1, this is also the highest
possible value for the classical mutual information, and as argued
above, there is no gap between this maximum and the actually
achievable classical information. The maximal information is retrieved
when Alice and Bob agree to use the Schmidt-bases for local projective
measurements on their respective parts of the quantum system. When
written in these bases, their measurements correspond to operators
represented by diagonal matrices, which clearly commute, and hence
there is no complementary issue in this case.

If a separable state $\rho_{AB}$ can be written as a convex sum of the
form
\bgeq
\rho_{AB}= \sum_{ij} p_{ij} \ket{\psi_i}_A \bra{\psi_i} \otimes
\ket{\phi_j}_B \bra{\phi_j} ,  \label{biorthogonalstates}
\edeq
with $\{ \ket{\psi_i}_A \}$ and $\{
\ket{\phi_j}_B \}$ fixed orthogonal bases of A and B respectively, the
maximal classical information is obtained if Alice measures in the $\{
\ket{\psi_i}_A \}$ basis and Bob measures in the $\{ \ket{\phi_j}_B
\}$ basis, respectively. There is no issue of complementarity, and
there is again no gap between the correlations in the state and the
classical mutual information.
In \cite{PHH08} it is shown, indeed, that this is the only kind of
states where $Q$ vanishes.

Let us now assume another separable state written on the form,
$\rho_{AB} = \sum_i p_i \ket{i} \bra{i} \otimes \rho_i^B$, where
$\rho_i^B$ denote density matrices for Bob's part of the quantum
system. Even though this is a separable state, that could have been
prepared by classical communication and local operations, we here
observe a consequence of complementarity: if Alice performs a
measurement in the $\ket{i}$-basis and finds the state
$\ket{i_0}$, Bob is in possession of the mixed state
$\rho_{i_0}^B$. The measurement yielding the maximum information to
Bob is then a projective measurement on the eigenbasis of
$\rho_{i_0}^B$. If Bob does not know $i_0$, he does not know on which
eigenbasis (for which $\rho_i^B$), he should perform his measurement,
and complementarity prevents him from obtaining such measurement
results for several bases simultaneously. Only if all $\rho_i^B$
commute, they are diagonal in the same basis, and the optimum
observables commute and are not complementary. In this limit, the gap
between the correlations in the state and the classical information
precisely vanishes.

\subsection{Non-classicality and the classical correlation locked in a quantum state} \label{sectqle}

It is shown in Ref.~\cite{DHLST04} that some hidden classical correlation
that cannot be obtained by local measurements can, nevertheless, be
``unlocked" with the help of a small amount of classical
communication (see Fig.~\ref{fig:protocols}(d)).

As an example, the following bipartite state is considered.
\bgeq
\rho_{AB} = \frac{1}{2d} \sum_{t=0}^{1} \sum_{i=0}^{d-1} \ket{t}_{A_1}
\bra{t}\otimes \ket{i}_{A_2} \bra{i} \otimes U_t \ket{i}_B \bra{i}
U_t^{\dagger}
\edeq
Here $U_0 =I$ is the identity operator on
$d$-dimensional Hilbert space, and $U_1$ is a unitary transformation
on the $d$-dimensional Hilbert space such that $\{ U_1 \ket{i} \}$ is
a mutually unbiased basis with respect to $\{\ket{i}\}$.

It is easily seen that $\rho_A = \frac{1}{2d} I$, $\rho_B =
\frac{1}{d} I$, and we have $S(A)=1+\log_2 d = S(AB)$, $S(B)=\log_2
d$, $S(A:B) = \log_2 d$.  The maximal classical mutual information
that can be extracted by local measurements and no classical
communication is shown in \cite{DHLST04} to be $I_{max} = \frac{1}{2}
\log_2 d$.  If, however, Alice makes a projective measurement on $A_1$
on the $\{|t=0\rangle,|t=1\rangle\}$ basis and sends her measured
value of $t$ as one classical bit to Bob, then he will know which of
the two bases $\{\ket{i}_B\}$ and $\{U_1\ket{i}_B\}$ should be used
for projective measurements in order to obtain the maximal classical
mutual information. This quantity takes on the value $\log_2 d$
bits, and the net increase of the classical correlation is
$\frac{1}{2} \log_2 d$ bits.  Note that this effect is in accordance
with our complementarity gap discussion in the previous subsection:
until Bob receives the information about the value of $t$,
complementarity of projective measurements in the bases
$\{\ket{i}_B\}$ and $\{U_1\ket{i}_B\}$, renders his measurements
inefficient, while knowing the value of $t$, he can extract the full
quantum mutual information of the remaining quantum state.

The amount of classical correlation, unlocked by the classical
communication, is $L(\rho_{AB})= \frac{1}{2} \log_2 d $
bits and it coincides with our $Q$ defined in Eq.(4):  $Q(A:B)=S(A:B)-I_{max}\{A:B\}=\frac{1}{2} \log_2
d=L(\rho_{AB})$.

As another example, we consider the following separable state.  \bgeq
\sigma_{AB} = \frac{1}{2d} \sum_{t=0}^{1} \sum_{i=0}^{d-1}
\ket{t}_{A_1} \bra{t} \otimes \ket{i}_{A_2} \bra{i} \otimes
\ket{i\oplus t}_B \bra{i\oplus t} \edeq $\sigma_{AB}$ has the marginal
states, $\sigma_A = \frac{1}{2d} I$, $\sigma_B = \frac{1}{d} I$, and
it is easy to show that $S(A)=1+\log_2 d = S(AB)$, $S(B)=\log_2 d$,
$S(A:B) = \log_2 d$ and $I_{max}=\log_2 d$. Hence $Q=0$ for this
state.  If Alice measures $A_1$ and sends her value of $t$ by one
classical bit to Bob as in the previous example, then the maximal
classical mutual information that they can obtain by measurements is still only $\log_2 d$
bits. There is
no classical correlation locked in the state $\sigma_{AB}$, i.e.,
$L(\sigma_{AB})=Q=0$. In this case, the same basis should be used by
Bob irrespective of the value of $t$ measured by Alice, and hence
there is also no complementarity issue in this case.

Let $I_L(\rho_{AB})$ denote the obtainable classical
correlation, maximized  for a given state, $\rho_{AB}$, over all
LOCC protocols, $\Pi$, while subtracting the cost, $c$, of the classical communication within the given protocol, $I_L(\rho_{AB}) \equiv max_{\Pi}\{ I'_{max} - c
\}$. \textit{I.e.}, $I'_{max}$ denotes the maximal amount of classical correlation
obtainable by the classical communication and by local measurements after exchange of $c$ bits of classical
communication.

From our definition of the maximal classical mutual information without any communication it follows that,
$I_{max} \leq I_L(\rho_{AB})$, and we obtain the maximal
locking effect of the quantum state $\rho_{AB}$, $L_{max} = I_L(\rho_{AB})- I_{max}$.

The examples in this section suggest a connection between $Q$ and the
locking effect, especially for states where $S(A:B)$ is not larger
than $S(A)$ and $S(B)$ (for example, separable states). Further studies may assess whether all or part of the information quantified
by $Q$ can be unlocked by transmission of classical information between
Alice and Bob. It is proposed in \cite{DG08} that the measurement-induced disturbance measure $D$
(see its definition in the next section) is connected to the
the locking effect, here we propose to use $Q$ instead of $D$ since $Q(\rho_{AB})$ is uniquely defined for any $\rho_{AB}$
and is a continuous function of $\rho_{AB}$ in contrast to the properties of $D$ (see Example C in Sec. V).

\section{Other measures of nonclassical correlations}

To remove the need to optimize the information over different measurements, it has been proposed \cite{Luo08}
to use a particular choice for the measurements. $I_e (\rho_{AB})$ thus denotes the classical correlation obtained by local projective measurements onto the eigenbases of the reduced density matrices on both Alice's and Bob's subsystem. The difference between
$S(A:B)$ and $I_e (\rho_{AB})$ is sometimes referred to as the measurement-induced disturbance measure $D(\rho_{AB})$.
It is obvious that $I_e (\rho_{AB}) \leq I_{max}(\rho_{AB})$, and hence, $D(\rho_{AB}) \geq Q(A:B)$.

An alternative measure, $C_A(\rho_{AB})$, of the classical correlation in a given state $\rho_{AB}$ is proposed in \cite{HV01},
\bgeq
C_A(\rho_{AB}) = max_{\{M_i^A\}} \{ S(\rho_B) - \sum_i p_i^A S(\rho_i^B) \} = S(\rho_B) - min_{\{M_i^A\}} \sum_i p_i^A S(\rho_i^B)
\label{defC1}
\edeq
where $\{M_i^A = K_i^{A \dagger} K_i^A \}$ is the set of POVM performed on system A and
$\rho_i^B = Tr_A (K_i^A \rho_{AB} K_i^{A\dagger})/ p_i^A $ is the state
of B conditioned on the outcome of the $i^{th}$ POVM on A, which occurs with the probability
$p_i^A  = Tr_{AB} (K_i^A \rho_{AB} K_i^{A\dagger})$.

The quantum discord $\mathcal{J}_{A}$ is now defined as the difference between the quantum mutual information
and $C_A(\rho_{AB})$,
\bgeq
\mathcal{J}_{A} = S(A:B) -C_{A}(\rho_{AB})
\label{discorddef1}
\edeq
Equivalently, we define $C_B(\rho_{AB}) = max_{\{M_s^B\}} \{ S(\rho_A) - \sum_s p_s^B S(\rho_s^A) \}$,
where the measurement is instead assumed to take place on system B, and the quantum discord $\mathcal{J}_{B} = S(A:B) -C_{B}(\rho_{AB})$.  The two discords $\mathcal{J}_{A,B}$ may not
in general be the same.

The quantum discord was originally defined in a similar manner in \cite{OZ01} but
with a restriction to projective measurements on subsystem A or B and without
the maximization over all local measurement, i.e., in a manner that depends explicitly on
the measurement performed on the subsystems. It is, indeed, difficult to determine (\ref{defC1}),
and in many studies (for example, references \cite{DSC08,DG08,Luo082})
only projective measurements are considered, and we hence denote
$C_{A(B)}^p(\rho_{AB})$ the classical correlation (\ref{defC1})
with maximization over projective measurements only,
and the corresponding quantum discord (\ref{discorddef1}) is denoted by $\mathcal{J}_{A(B)}^p$.
It is clear that $C_{A(B)}^p(\rho_{AB}) \leq C_{A(B)}(\rho_{AB})$,
and $\mathcal{J}_{A(B)}^p \geq \mathcal{J}_{A(B)}$,
but it is not clear whether they actually coincide.

Since $C_{A(B)}(\rho_{AB})$ can be viewed as the Holevo bound on the accessible
information Bob (Alice) can obtain by his (her) local
measurement on the ensemble $\{p_i^A, \rho_i^B\}$
($\{p_s^B, \rho_s^A\}$),
it follows that  $I_{max} \leq C_{A(B)}(\rho_{AB})$ and $I_{max}^p \leq C_{A(B)}^p(\rho_{AB})$,
where we have introduced a similar superscript $p$ for the maximum mutual information obtainable with restriction
to projective measurements by Alice and Bob.
Therefore, it follows that our measure of nonclassicality is larger than or equal to the quantum discord,
both when these quantities are defined with respect to general measurements,
$Q(A:B) \geq \mathcal{J}_{A(B)}$, and with respect to projective measurements, $Q^p(A:B) \geq \mathcal{J}_{A(B)}^p$.

\section{Quantitative examples.}

The purpose of this section is to study some examples to illustrate the practical difficulties of dealing with the different classical information measures and to display the non-trivial interplay between the different aspects of nonclassical correlations of quantum states.

\subsection{Example: Maximum classical mutual information by projective measurements on qubits.}

Consider a family of two-qubit states, where the reduced density
matrices of both qubits are proportional to the identity operator. Such
states can be written in terms of Pauli matrices,
\bgeq \rho_{AB} =\frac{1}{4} \left( I\otimes I +
  \sum_{j,k=1}^3 w_{jk} \sigma_j \otimes \sigma_k \right)
\label{eg2qubitstate}
\edeq
and can  be transformed by a local unitary transformation to
the following form
\bgeq \sigma_{AB} =\frac{1}{4} \left( I\otimes I +
  \sum_{j=1}^3 r_{j} \sigma_j \otimes \sigma_j \right) \edeq where
$|r_j|$ are the singular values of the matrix $w_{jk}$. The eigenvalues
of $\sigma_{AB}$ or $\rho_{AB}$ are given by $\lambda_0=\frac{1- r_1-
  r_2 - r_3}{4}$, $\lambda_1=\frac{1- r_1+ r_2 + r_3}{4}$,
$\lambda_2=\frac{1+r_1-r_2 +r_3}{4}$, and $\lambda_3=\frac{1+r_1+r_2
  -r_3}{4}$.  Therefore
\begin{equation}
  S(A:B)
  =
  2 -H\{ \lambda_0, \lambda_1,\lambda_2, \lambda_3 \}
  = 2 + \sum_{j=0}^3 \lambda_j \log_2 \lambda_j.
\end{equation}

Local projective measurements performed on A and B can be written
$M_{\pm}^A =\frac{1}{2} (I \pm \overrightarrow{n_A} \cdot
\overrightarrow{\sigma_A})$ and $M_{\pm}^B =\frac{1}{2} (I \pm
\overrightarrow{n_B} \cdot \overrightarrow{\sigma_B})$,
parametrized by unit vectors
$\overrightarrow{n_A}$ and $\overrightarrow{n_B}$, respectively.  The probabilities
to obtain the results $++$, $+-$, $-+$ and $--$ are
$\frac{1+\delta}{4}$, $\frac{1-\delta}{4}$, $\frac{1-\delta}{4}$ and
$\frac{1+\delta}{4}$, with $\delta = r_1 n_{Ax} n_{Bx} + r_2 n_{Ay}
n_{By} +r_3 n_{Az} n_{Bz}$, and the mutual information of the measurement
results is given by
\bgeq
I\{A:B\} = 1- H\{\frac{1+\delta}{2},
\frac{1-\delta}{2}\} .
\edeq
The maximal mutual
information is obtained when $|\delta|$ reaches its maximum.  Let
$n_A^T =(n_{Ax}, n_{Ay}, n_{Az})$ (similarly for B) denote a row
vector and $D=diag\{r_1,r_2,r_3\}$ denote a diagonal matrix. We then have
\bgeq
|\delta| = |n_A^T D n_B| \leq \frac{n_B^T D^2 n_B}{\parallel D n_B \parallel} =
\sqrt{n_B^T D^2 n_B} \leq r_m
\edeq
where $r_m = max \{ |r_1|, |r_2|,
|r_3|\}$ determined by the singular values of the matrix $w_{jk}$ in
(\ref{eg2qubitstate}). Therefore the maximal obtainable classical
correlation of the family of two-qubit states in (\ref{eg2qubitstate})
by local projective measurements is given by \bgeq I_{max}^p = 1-
H\left\{\frac{1+r_m}{2}, \frac{1-r_m}{2}\right\}.  \edeq This value is achieved
when A and B are projected onto the local bases that give the singular
value decomposition of $w_{jk}$.

With the restriction to projective measurements we thus find
\begin{equation}
  \label{eq:ex1_Qp}
  Q^p(A:B)
  =
  S(A:B)- I^p_{max}
  =
  1
  +
  H\left\{\frac{1+r_m}{2}, \frac{1-r_m}{2}\right\}
  -
  H\left\{\lambda_0,\lambda_1,\lambda_2,\lambda_3\right\},
\end{equation}
which happens to be equal to the quantum
discord $\mathcal{J}_{A(B)}^p$ obtained in \cite{Luo082}. As noted in \cite{Luo082},
$\mathcal{J}_{A(B)}^p$, and hence $Q^p(A:B)$, are larger than the entanglement of
formation for some states and smaller for others.

\subsection{ Example:  Werner states in arbitrary dimensions,
non-classicality and entanglement.}

Even though numerous studies suggest a strong relation between
nonclassical correlations and quantum entanglement, $Q$ is not a measure of entanglement since
separable quantum states exist, for which the maximum classical mutual
information obtained by measurements does not exhaust the quantum
mutual information.

As an example, we consider a Werner states of a $d\times d$ dimensional system \cite{Wer89},
\bgeq
\rho_{AB}=(I-\alpha P) / (d^2 -d\alpha)   \label{wernerstate}
\edeq
where
$P=\sum_{i,j=1}^d \ket{i} \bra{j} \otimes \ket{j}\bra{i}$.

For this state, a straightforward calculation yields
\begin{equation}
  S(A:B)
  =
  2 \log_2 d
  +\frac{(1+\alpha)(d-1)}{2(d-\alpha)}
  \log_2\frac{1+\alpha}{d(d-\alpha)}
  +
  \frac{(1-\alpha)(d+1)}{2(d-\alpha)}
  \log_2 \frac{1-\alpha}{d(d-\alpha)}
  ,
\end{equation}
while projective measurements in the bases $\{|i\rangle,|j\rangle\}$, used in the definition of $P$, yield the maximal mutual information
\begin{equation}
  I^p_{max}
  =
  \log_2 (\frac{d}{d-\alpha})
  + \frac{1-\alpha}{d-\alpha} \log_2 (1-\alpha)
  ,
\end{equation}

\begin{figure}
\begin{center}
\includegraphics[angle=0,height=1.5in]{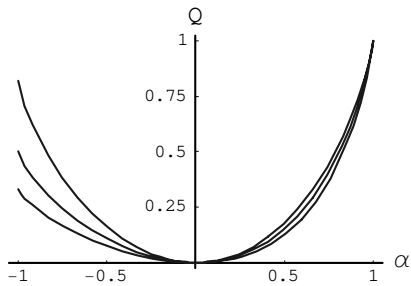}
\end{center}
\caption{Plot of Q for Werner states with $d=2$, $3$, and $10$ (from below).}
\label{figwerner}
\end{figure}

$Q^p(A:B)=S(A:B)-I^p_{max}$ is shown, for $d=2,3,10$ in Fig. \ref{figwerner}. We see that $Q^p (A:B) = 0 $
only when $\alpha=0$, i.e., when $\rho_{AB}$ is the identity matrix, which is itself a product state of subsystem identity density matrices.
For all other values of $\alpha$,  $Q^p(A:B) \neq 0 $ even when $\rho_{AB}$ is a separable state($\alpha \leq 1/d$).

\subsection{Example: Non-uniqueness and discontinuous behavior of $I_e$ and $D$ versus the uniqueness and
continuous behavior of $I_{max}$ and $Q$.}

The Werner state (\ref{wernerstate}) may also be used to illustrate another property of
the classical information measures based on projective measurements. The local density
matrices $\rho_A$ and $\rho_B$ of the Werner state are both proportional to the identity
matrices. This implies that the choice of projective measurements in the eigenbases of
$\rho_A$ and $\rho_B$ is not uniquely defined, and may in fact provide a range of values
for the classical mutual information of the detection records. Maximum information is
obtained, as argued in the previous example, by the use of the same bases
$\{|i\rangle,|j\rangle\}$ on the two systems as are used in the definition of the
operator $P$. Using that pair of eigenbases, $I_e=I^p_{max}$. If on the other hand,
Alice uses the basis $\{|i\rangle\}$, and Bob uses an eigenbasis $\{|j_u\rangle\}$,
which is mutually unbiased \cite{Ivanovic81,wootters89}, to the basis  $\{|j\rangle\}$, we obtain $I_e=0$.
This reflects the complementarity between the information available by measurements
on mutually unbiased bases (see Sec. \ref{sectcompmub} and Refs. \cite{kraus87, MU88, hall95, hall97, wym08, wym082}),
but it also shows that $I_e$ is not uniquely defined, when the local density matrices have degenerate eigenvalues.

Another consequence of this ambiguity is that $I_e$ and hence $D$ become non-continuous functions of the density matrix in the vicinity of these degeneracies. We may for example add an infinitesimal term, $\epsilon \left( \sum_{i=1}^d \lambda_i \ket{i} \bra{i} \otimes
\sum_{j=1}^d \delta_j \ket{j} \bra{j} \right)$ to the Werner state, with different $\lambda_i$'s and $\delta_j$'s, to ensure that for any nonzero $\epsilon$, the standard bases are the unique eigenstates of $\rho_A$ and $\rho_B$. In contrast, adding a similar infinitesimal term $\epsilon \left( \sum_{i=1}^d \lambda_i \ket{i} \bra{i} \otimes
\sum_{j=1}^d \delta_j \ket{j_u} \bra{j_u} \right)$, involving the mutually unbiased basis of system B, causes the other unique choice of basis. These two choices have infinitesimally close density matrices, but they have unique values of  $I_e \approx \log_2 (\frac{d}{d-\alpha})+ \frac{1-\alpha}{d-\alpha} \log_2 (1-\alpha)$ and $I_e=0$, respectively.

In contrast, $I_{max}$ and $Q(\rho_{AB})$ (as well as $I_{max}^p$ and $Q^p(\rho_{AB})$) do not have such a problem,
they are uniquely defined for any $\rho_{AB}$ and are continuous functions of $\rho_{AB}$.

\subsection{Example: States with finite Q and vanishing quantum discord $\mathcal{J}_{A}$ or $\mathcal{J}_{B}$.}

Since quantum discord and our non-classicality $Q$ both try to quantify the non-classical correlation in a quantum state,
it is very interesting to know the difference between these two quantities.
As shown in Sec. IV, our non-classicality $Q$ is always no less than quantum discord $\mathcal{J}_{A(B)}$,
i.e., $Q(A:B) \geq \mathcal{J}_{A(B)}$ (and $Q^p(A:B) \geq \mathcal{J}_{A(B)}^p$) for any state $\rho_{AB}$.
When $\rho_{AB}$ is a pure state, it is easy to verify that $Q(A:B) = \mathcal{J}_{A(B)}= S(A)$, namely, both
our non-classicality $Q$ and quantum discord $\mathcal{J}_{A(B)}$ are equal to the amount of entanglement in the state.
And when $\rho_{AB}$ has the form in (\ref{biorthogonalstates}), it is also obvious to show that
both non-classicality $Q$ and quantum discord $\mathcal{J}_{A(B)}$ vanish.

Now we show that $Q$ is finite while $\mathcal{J}_{A}=0$ for a family of states.
It is shown in \cite{HV01} that for the family of states,
\bgeq
\rho_{AB} = \sum_i p_i \ket{i}_A \bra{i} \otimes \rho_i^B
\label{goodex}
\edeq
where $\{\ket{i}\}$ is a set of orthonormal states of A,
\bgeq
C_A(\rho_{AB}) = C_A^p(\rho_{AB})= S(\rho_B) - \sum_i p_i S(\rho_i^B) = S(A:B).
\edeq
It therefore follows that the quantum discord, $\mathcal{J}_{A} =\mathcal{J}_{A}^p=0$ for this family of states.
And it is also shown in \cite{Datta08thesis} that this is the only kind of states with a vanishing $\mathcal{J}_{A}$.
This is also precisely the kind of states discussed at the end of Sec. III A, illustrating the role of complementarity,
when Bob attempts to extract maximal information by measurements but is faced with the problem,
that the different $\rho_i^B$ will generally require complementary optimal measurement strategies.
If the different $\rho_i^B$s do not commute, $Q>0$ for the states in (\ref{goodex}), and the classical information
cannot exhaust the quantum mutual information.
The states in (\ref{goodex}) with non-commuting $\rho_i^B$s are the only kind of states with a finite $Q$ and
vanishing $\mathcal{J}_{A}$.

However, if both $\mathcal{J}_{A}=0$ and $\mathcal{J}_{B}=0$ for a state $\rho_{AB}$, then
$\rho_{AB}$ must have the form in (\ref{biorthogonalstates}), and therefore, $Q=0$. So there does not exist a state
such that $Q$ is finite while both $\mathcal{J}_{A}$ and $\mathcal{J}_{B}$ is zero.

\subsection{Example: $I^p_{max} < I_{max}$.}

It is generally difficult to find the optimum measurement strategy both on the general case of POVMs and when we are restricted to projective measurements. Even in the case where $\rho_{AB}$ is a state of two qubits, it is not clear whether local projective measurements
are sufficient to extract the maximal classical correlation, i.e., whether
$I_{max}(\rho_{AB}) = I_{max}^p(\rho_{AB}) $.
One attempt to address this issue was given in \cite{TP08}, in connection with a different problem:
a quantum binary channel is used to communicate symbols encoded in two states,
and the mutual information is maximized over the receiver's measurement strategies.
It is shown in \cite{TP08} that projective measurements yield the same information as two-outcome POVMs. Two-outcome measurements, however, do not exhaust all POVMs, and the analysis in \cite{TP08}, does not rule out the possibility that, e.g., three-outcome POVMs may lead to a higher $I_{max}$ than projective measurements on the qubits.
By contrast, it is shown in \cite{Hol73-2} that when symbols are encoded in three states,
an appropriate POVM measurement yields strictly more information than any projective measurement.

Based on the analysis in \cite{Hol73-2}, we can construct an example where we can prove that $I^p_{max} < I_{max}$.
Suppose system A is a qutrit and system B is a qubit, and
\bgeq
\rho_{AB} = \sum_{i=1}^3 \frac{1}{3} \ket{i}_A \bra{i} \otimes \ket{\phi_i}_B\bra{\phi_i}
\label{goodex5}
\edeq
with the Bloch vectors of the pure states $\ket{\phi_i}_B$ forming equal angles $\frac{2\pi}{3}$ in the same plane.
In order to extract the maximal mutual information, Alice simply projects system A onto
the basis states $\{ \ket{i}_A  \}$ (see below). Now system B is in one of the three state $\ket{\phi_i}_B$ with an equal
probability of $1/3$, and
Bob needs to perform an appropriate measurement in order to extract the maximal mutual information.
This is exactly the situation in \cite{Hol73-2}
where a certain POVM measurement with 3 elements
extracts strictly more information
than any projective measurement on Bob's qubit.
Therefore, for the state in (\ref{goodex5}), $I^p_{max} < I_{max}$.

We need to show that for the state in (\ref{goodex5}), and more generally for the states in (\ref{goodex}),
in order to obtain
$I_{max}$, Alice should indeed simply project system A onto the basis states $\{ \ket{i}_A  \}$.
Assume that the best measurement strategy for Bob is a rank-one POVM $\{ \sum_s \ket{K_s^B} \bra{K_s^B} =I\}$,
then after Bob obtains result $s$, the state of system A is
$\rho_s^A=\sum_i |\bra{K_s^B} \rho_i^B \ket{K_s^B}|^2 p_i \ket{i} \bra{i} $ (not normalized).
If Alice performs a POVM $\{ \sum_j \ket{K_j^A} \bra{K_j^A} =I\}$, the joint probability that
she gets $j$ and Bob gets $s$ is given by
$p_{js} = \sum_i p_i |\bra{K_s^B} \rho_i^B \ket{K_s^B}|^2  |\left\langle K_j^A|i\right\rangle |^2$.
However, the same joint probability distribution can also be obtained if Alice first projects system A onto
the basis states $\{ \ket{i}_A  \}$, and thereafter
she performs the POVM $\{ \sum_j \ket{K_j^A} \bra{K_j^A} =I\}$.
This is a Markov chain, as $p(j|is)=p(j|i)=|\left\langle K_j^A|i\right\rangle |^2$; and
we have the same joint probability distribution $p_{js}$.
Therefore the mutual information $I\{A:B\}$ obtained from the joint probability distribution $\{ p_{is}\}$
is larger or equal to the one obtained from $\{ p_{js}\}$.
In other words, the maximal mutual information can always be obtained when
Alice projects her system onto the basis states $\{ \ket{i}_A  \}$ and Bob chooses an appropriate measurement strategy.

\section{Complementarity of classical correlation with different MUBs on one side}
\label{sectcompmub}

In this section we will quantify the consequences of complementarity by a specific analysis of
projective measurements in different mutually unbiased bases, i.e., bases,
where each basis vector in one basis has the same squared overlap with all basis vectors in the other bases.
In the following, we suppose Bob performs a fixed POVM measurement, while Alice performs
a projective measurement along one of two or more MUBs at her choice.  We shall derive upper bounds of the sum of
classical correlation with Alice's different choices of MUBs.

\begin{theorem}
Suppose Bob performs an arbitrary general measurement
$\{ M_s =K_s^{B\dagger} K_s^B | \sum_s M_s =I\}$ on B,
while Alice performs on A a complete projective measurement onto either
the basis
$\{ |i_1> | i=1,\cdots,d_A \}$ (hence define $I_1 \{A:B\}$),
or on a second basis  $\{ |i_2> | i=1,\cdots,d_A \}$ ($I_2 \{A:B\}$) which is mutually unbiased to the first basis,
then
\bgeq
I_1\{A:B\} +I_2 \{A:B\} \leq  \log_2 d_A.
\edeq
\end{theorem}

{\bf Proof.} We have
\bgeq
I_1\{A:B\} +I_2 \{A:B\} =  H\{ p^{(1)}_i;i \} + H\{ p^{(2)}_i;i \} -
\sum_s p(s) \left( H\{ p^{(1)}_{i|s};i \} +  H\{ p^{(2)}_{i|s};i \} \right) .
\edeq
Here $p(s) =Tr (M_s^B \rho_B)$ and $p^{(m)}_{i|s} =\bra{i_m} \rho_s^A \ket{i_m}$ ($m=1,2$),
with $\rho_s^A$ as the normalized state of A conditional on B's result $s$,
i.e., $\rho_s^A = Tr_B (K_s^B \rho_{AB} K_s^{B\dagger}) / Tr_{AB}(K_s^B \rho_{AB} K_s^{B\dagger})$.
The entropic uncertainty relation \cite{MU88} implies
\bgeq
H\{ p^{(1)}_{i|s};i \} +  H\{ p^{(2)}_{i|s};i \} \geq \log_2 d_A .
\edeq
Therefore
\bgeq
I_1\{A:B\} +I_2 \{A:B\} \leq  H\{ p^{(1)}_i;i \} + H\{ p^{(2)}_i;i \} - \log_2 d_A .
\edeq
Together with the fact that $H\{ p^{(m)}_i;i \}\leq \log_2 d_A$ ($m=1,2$), this implies the proposition.

The number of mutually unbiased bases in a Hilbert space of dimension $d$ is not generally known,
but for $d\geq 2$, there are at least 3 such bases, and for $d$ a power of a prime, there are $d+1$ MUBs. Given the existence of $M$ MUBs, we can define
the local-measurement-induced mutual information, $I_m $, when Alice projects her
system onto the $m$th MUB while Bob performs the same fixed general measurement on
his system, and we can introduce the sum,
\bgeqn
I_{tot} \equiv \sum_{m=1}^{M} I_m &=& \sum_m H\{ p^{(m)}_i;i \} - \sum_m \sum_s p(s) H\{ p^{(m)}_{i|s};i \} \\
&=& \sum_m H\{ p^{(m)}_i;i \} - \sum_s p(s) \sum_m H\{ p^{(m)}_{i|s};i \} .
\label{MmubsinfoII}
\edeqn

For this we have the following
\begin{theorem}  \label{Mmubs1}
\bgeq
I_{tot}=\sum_{m=1}^{M} I_m \leq
M \log_2 \frac{d_A}{K +1}
+ K \left( (K +1) \frac{d_A+M-1}{d_A} -M \right)
\log_2 \left( 1+ \frac{1}{K} \right)
\leq
M \log_2 \frac{d_A+M-1}{M}  .
\label{Mmubs10}
\edeq
with $K=\lfloor \frac{Md_A}{d_A+M-1} \rfloor$, and
\bgeq
I_{tot}=\sum_{m=1}^{M} I_m \leq \frac{M}{2} \log_2 d_A .  \label{Mmubs1001}
\edeq
Here (\ref{Mmubs10}) is stronger than (\ref{Mmubs1001}) when $M>\sqrt{d}+1$
and weaker than (\ref{Mmubs1001}) when $M < \sqrt{d}+1$.
\end{theorem}
{\bf Proof.} Since $\sum_{m=1}^M H\{ p^{(m)}_i;i \} \leq M \log_2 d_A$,
(\ref{Mmubs10}) follows from (\ref{MmubsinfoII}) and the entropic uncertainty relation
\cite{wym08}
\bgeqn
\sum_{m=1}^M H\{ p^{(m)}_{i|s};i \} \geq
M \log_2 (K +1)
- K \left( (K +1) \frac{d_A+M-1}{d_A} -M \right)
\log_2 \left( 1+ \frac{1}{K} \right) \geq
M \log_2 \frac{Md_A}{d_A+M-1} ,  \label{Mmubeq11}
\edeqn
with $K=\lfloor \frac{Md_A}{d_A +M-1} \rfloor$; and
(\ref{Mmubs1001}) follows from (\ref{MmubsinfoII}) and the inequality \cite{BW07,wym08}
\bgeq
\sum_{m=1}^M H\{ p^{(m)}_{i|s};i \} \geq \frac{M}{2} \log_2 d_A .
\edeq

Proposition \ref{Mmubs1} certainly holds true when $d_A$ is a power of a prime and $M=d_A+1$. However using the
results from \cite{sanchez-ruiz95}
\bgeq
\sum_{m=1}^3 H\{ p^{(m)}_i;i \} \leq  3 H\{ \frac{1+R}{2},  \frac{1-R}{2}\}  \textrm{ when $d_A=2$}
\edeq
where $R=\sqrt{\frac{2 Tr \rho_A^2-1}{3}}$, and
\bgeq
\sum_{m=1}^{d_A+1} H\{ p^{(m)}_i;i \} \leq (d_A+1) \log_2 d_A - \frac{(d_A-1)(d_A Tr(\rho_A^2) -1) \log_2 (d_A-1)}{d_A(d_A-2)}
\textrm{ when $d>2$,} \label{1333}
\edeq
together with (\ref{Mmubeq11}) this leads to the following tighter bound
\begin{theorem}  \label{themub1}
For $d_A=2$ and $M=3$ we have
\bgeq
I_{tot} \leq 3 H\{ \frac{1+R}{2},  \frac{1-R}{2}\} -2  \label{qubit3mub11}
\edeq
where $R=\sqrt{\frac{2 Tr \rho_A^2-1}{3}}$, and when $d_A$ is a power of a prime and $M=d_A+1$ we have
\bgeqn
I_{tot} \leq     - \frac{(d_A-1)(d_A Tr(\rho_A^2) -1) \log_2 (d_A-1)}{d_A(d_A-2)}
  + \left\{
\begin{array}{ll}
(d_A+1) \log_2 (\frac{2d_A}{d_A+1}) &  \textrm{if $d_A$ is odd,} \\
d_A+1 + (\frac{d_A}{2}+1) \log_2 (\frac{d_A}{d_A+2}) & \textrm{if $d_A$ is even.}
\end{array}
\right.  \label{eqdplus1mubstr}
\edeqn
\end{theorem}

We have the following upper bounds which are independent of the state.
\begin{corollary} \label{cormub11}
When $d_A$ is a power of a prime, and $M=d_A+1$,
\bgeqn
I_{tot} =\sum_{m=1}^{d_A+1} I_m &\leq&
   \left\{
\begin{array}{ll}
(d_A+1) \log_2 (\frac{2d_A}{d_A+1}) &  \textrm{when $d_A$ is odd,} \\
d_A+1 + (\frac{d_A}{2}+1) \log_2 (\frac{d_A}{d_A+2}) & \textrm{when $d_A$ is even.}
\end{array}
\right.   \label{eqamub1}    \\
I_{tot} =\sum_{m=1}^{d_A+1} I_m &<& d_A \label{eqamub12}
\edeqn
\end{corollary}

The inequality in (\ref{eqamub1}) follows from (\ref{eqdplus1mubstr}) with the
observation that $Tr(\rho_A^2)\geq 1/d_A$. By expansion of the logarithm, we can verify that
the quantity on the right hand side of (\ref{eqamub1}) is a number that lies between $d_A-1$ and $d_A$,
but strictly less than $d_A$ for any $d_A\geq 2$, hence we have (\ref{eqamub12}).

It should be pointed out that, for a special case when the shared state $\rho_{AB}$
is a maximally entangled state, the problem considered in this section is equivalent
to the problems considered in \cite{hall97,wym082} according to the {\sl source duality} \cite{hall95} or
the atemporal diagram approach \cite{griffiths05}.

\section{Nonclassical correlations in quantum computation}

The previous sections have all been of a formal nature and have presented examples of quantitative differences between the magnitudes of different correlation measures. In this section, we consider states whose nonclassicality is linked with their performance as resources for a specific task:
speed-up of quantum computing. Entanglement is an interesting and valuable quantum resource in quantum communication and computing. But it is not indispensable, as illustrated by the paradigmatic, deterministic quantum computation with one
quantum bit (DQC1) proposed in \cite{KL98}. This example provides an exponential speed-up over the best known classical
algorithms and yet has a limited amount of entanglement and, in some regimes, no distillable entanglement
at all \cite{DFC05}.  We shall see that the non-classicality Q defined above is closely related to the non-local resource responsible for the speed-up of the computation in the DQC1 model.

\begin{figure}
\begin{center}
\includegraphics[angle=0,height=1in]{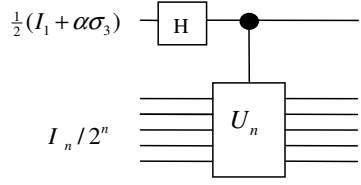}
\end{center}
\caption{The DQC1 model.}
\label{figdqccircuit}
\end{figure}

In the generalized DQC1 model \cite{DFC05}, one attempts to estimate the normalized trace, $2^{-n} \textrm{Tr}(U_n)$,
of a unitary operator $U_n$ acting on $n$ qubits. One assumes that the $n$ qubits are all prepared in the fully mixed
state with equal probabilities on both qubit states $|0\rangle$ and $|1\rangle$, while the application of the unitary
operator $U_n$ is controlled by a single qubit, in the polarized initial state $\frac{1}{2} (I_1 + \alpha \sigma_3)$.
The overall state of the $n+1$ qubits after the Hadamard transformation on the control qubit and the interaction,
illustrated in Fig. \ref{figdqccircuit} is given by
\bgeq
\label{dqc1}
\rho_{n+1} (\alpha) =2^{-(n+1)} \{ \ket{0} \bra{0} \otimes I_n + \ket{1} \bra{1} \otimes I_n +
\ket{0} \bra{1} \otimes \alpha U_n^\dagger +\ket{1} \bra{0} \otimes \alpha U_n
\}
\edeq
where $I_n$ and $U_n$ are, respectively, the identity and unitary operator on the Hilbert space of $n$ qubits.
This state is separable  with respect to the division between the control qubit and the collection of target qubits, which is most readily seen by expanding the state on the eigenstate basis
$\{|e_i\rangle\}$ of the unitary operator $U_n$ with eigenvalues $\exp(i\theta_i)$,
\bgeq
\label{dqc11}
\rho_{n+1} (\alpha) = 2^{-(n+1)}  \sum_i (\ket{0} \bra{0} + \ket{1} \bra{1} +
\alpha e^{-i\theta_i} \ket{0} \bra{1}  +\alpha e^{i\theta_i}\ket{1} \bra{0})\otimes |e_i\rangle \langle e_i|.
\edeq
This equation shows how measurement of the expectation value of the $\sigma_1$ and $\sigma_2$ Pauli matrices
on the single control qubit give respectively $\alpha 2^{-n}\sum_i \cos(\theta_i)$ and  $\alpha 2^{-n}\sum_i \sin(\theta_i)$
and hence provides the normalized trace of $U_n$, and
the number of runs required to achieve a given precision is proportional to $ 1/ \alpha^2$ but independent of the dimension $2^n$ of the $n$-qubit Hilbert space.
The control qubit (system A) is completely disentangled from the $n$-qubit target register (system B), and the exponential speed-up of the DQC1 model over classical algorithms is hence not due to entanglement. Eq. (\ref{dqc11}) is precisely of the form (\ref{goodex}) and it hence has $Q> 0$, unless all eigenvalues $\exp(i\theta_i)$ are identical, and $U$ is the identity operator (no action on the target registers).
It is straightforward to calculate the quantum mutual information
\bgeq
S(A:B)= H\{\frac{1+|\beta|}{2}, \frac{1-|\beta|}{2}\} - H\{\frac{1+\alpha}{2}, \frac{1-\alpha}{2}\}
\rightarrow 1- H\{\frac{1+\alpha}{2}, \frac{1-\alpha}{2}\},
\label{sdq1}
\edeq
where $\beta=2^{-n}\alpha \textrm{Tr}(U_n) \rightarrow 0$ for a typical unitary $U_n$. By typical, we mean it is chosen randomly
according to the Haar measure on $\mathcal{U}(2^n)$. For such a unitary, the eigenvalues are almost uniformly
distributed on the unit circle with large probability \cite{Diaconis03}.

Now we proceed to calculate the classical mutual information. Suppose system A is projected onto two basis states
$\ket{1_A}=\{\cos \theta \ket{0} + e^{i \phi} \sin \theta \ket{1} \}$ and
$\ket{2_A}=\{\sin \theta \ket{0} - e^{i \phi} \cos \theta \ket{1} \}$, and system B is projected onto
a set of basis states $\ket{s_B}$. The joint probability $p_{is}$ for system A to be found in the $i$th state and system B
to be found in the $s$th is given by
\bgeq
p_{is} = 2^{-(n+1)} \{ 1- (-1)^i \alpha \sin 2\theta \cdot Re [ e^{-i \phi} \bra{s_B} U_n \ket{s_B} ] \}
\edeq
From this joint probability distribution, we determine the classical mutual information of the measurements,
and in order to calculate $I_{max}^p$, we need to maximize over all possible
choices of the local bases for both A and B. The maximization will depend on the specific unitary $U_n$, and is prohibitively complicated, and we will only consider typical unitaries and assume eigenvalues uniformly distributed on the unit sphere. We then get
\bgeq
I\{ A:B \} \approx 1- 2^{-n} \sum_{s=1}^{2^n} H\{ \frac{1+\delta_s}{2}, \frac{1-\delta_s}{2}\}
\edeq
with $\delta_s =\alpha \sin 2\theta Re[ e^{-i \phi} \bra{s_B} U_n \ket{s_B} ]$.
The maximal value of this $I\{ A:B \}$ is obtained when $\delta_s \rightarrow \alpha \cos \frac{2\pi s}{2^n}$,
which can be achieved when $\theta \rightarrow \pi/4$, $\phi \rightarrow 0$ and $\ket{s_B}$ are chosen as
the eigenstates of $U_n$.

Therefore, assuming also $\beta \rightarrow 0$ in the evaluation of $S(A:B)$ (\ref{sdq1}),  the nonclassicality is given by
\bgeq
Q \rightarrow 2^{-n} \sum_{s=1}^{2^n} H\{ \frac{1+\delta_s}{2}, \frac{1-\delta_s}{2}\} -
H\{\frac{1+\alpha}{2}, \frac{1-\alpha}{2}\}
\edeq
with $\delta_s = \alpha \cos \frac{2\pi s}{2^n}$.

\begin{figure}
\begin{center}
\includegraphics[angle=0,height=1.5in]{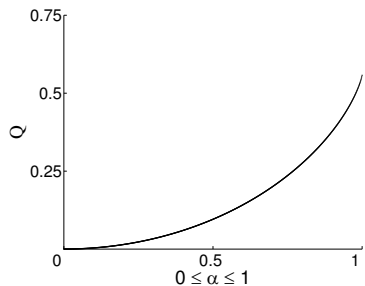}
\end{center}
\caption{Plot of Q of the DQC1 model for 10+1 qubits as a function of $\alpha$.}
\label{figdqc1n10}
\end{figure}

For $n=10$, $Q$ is shown in Fig. \ref{figdqc1n10} as a function of the control qubit polarization $\alpha$.
Even though there is no entanglement between the control qubit and the other qubits at
any point during the computation, $Q$ is nonzero and increases
as the polarization $\alpha$ increases and the DQC1 model becomes more effective.
This is suggestive that $Q$ quantifies the correlations that enable the advantage in quantum computation.

In \cite{DSC08}, the quantum discord, and in \cite{Luo08,DG08}, the measurement-induced disturbance measure have been similarly used  to characterize the correlations in the DQC1 model. The behavior of $Q$ for the DQC1 model
is qualitatively similar to that of the quantum discord and of the measurement-induced disturbance.

\section{Conclusion}

In this paper, we have investigated the quantum and classical correlations stored in bipartite quantum states. We have suggested to quantify the classical correlations by the maximum classical information, that can be retrieved by local measurements, and shown that the difference between the quantum mutual information of the state and this $I_{max}$ provides a good measure of non-classicality, in the sense that it vanishes when there are no non-classical correlations, and it gives non-vanishing results for states whose non-classicality are not revealed by other measures. We have argued that the state-dependent gap between the quantum and classical mutual information is associated with the complementarity between the local observables which together characterize the properties of the system. The same gap provides a quantitative measure of the computing power in the DQC1 proposal, characterized so far by other measures, and examples suggest that a non-vanishing value for the gap also witnesses a non-vanishing locking effect.

Our measure as well as some of the alternative measures of correlations suffer from the immense difficulty of
calculating their value, because they assume an optimization over all possible measurement strategies.
This implies, like in the theory of entanglement, that many scattered results exist, but we do not yet have
a full overview of the types of correlations that can be extracted from quantum states.
We hope that our work may stimulate the search for analytical and numerical methods for the effective
determination of these measures.
We also find it very interesting to investigate under which circumstances classical communication
in the presence of a complementarity gap may be used to partly or fully unlock the classical correlations.

\section*{Acknowledgments}

The authors wish to thank Sixia Yu for helpful discussions.
S. W. also wishes to acknowledge support from the NNSF of China (Grant No. 10604051), the CAS, and the National Fundamental Research Program.

\section*{Appendix}

{\bf Proof of proposition \ref{the1}.} The first part of the proof is
similar to that of the Holevo bound \cite{Hol73, FC94, YO93}.
Suppose Alice introduces an ancilla $A_1$ and Bob introduces $B_1$,
the systems $A_1$, $A$, $B$, $B_1$ are in the following initial state,
\bgeq \rho_{A_1 A B B_1} = \ket{0}_{A_1} \bra{0} \otimes \rho_{AB}
\otimes \ket{0}_{B_1} \bra{0} .  \edeq Now Alice performs the
measurement $\{ M_i^A \}$ on $A$ and stores her measurement result in
$A_1$, the overall state becomes, \bgeq \rho_{A'_1 A' B B_1} =\sum_i
\ket{i} \bra{i} \otimes K_i^{A} \rho_{AB} K_i^{A\dagger} \otimes
\ket{0}_{B_1} \bra{0} .  \edeq Here $K_i^{A\dagger} K_i^A = M_i^A$,
and $A'_1$, $A'$ denote systems $A_1$ and $A$ after Alice's
measurement. Then Bob performs his measurement $\{ K_s^{B\dagger}
K_s^B = M_s^B\}$ on $B$ and stores his measurement result in $B_1$,
the overall state becomes \bgeqn \rho_{A'_1 A' B' B'_1} &=& \sum_{is}
\ket{i} \bra{i} \otimes
\left( K_i^{A} \otimes K_s^{B} \rho_{AB} K_i^{A\dagger} \otimes K_s^{B\dagger} \right) \otimes \ket{s} \bra{s} \nonumber \\
&=& \sum_{is} \ket{i} \bra{i} \otimes \left( p_{is} \rho_{is}^{AB}
\right) \otimes \ket{s} \bra{s} \edeqn where $p_{is}= Tr \left(
  K_i^{A} \otimes K_s^{B} \rho_{AB} K_i^{A\dagger} \otimes
  K_s^{B\dagger} \right)$ is the joint probability that Alice obtains
result $i$ and Bob obtains result $s$.  $B'_1$, $B'$ denote systems
$B_1$ and $B$ after Bob's measurement.  The mutual information can be
written as \bgeqn
I\{A:B\} &=   & S(A'_1:B'_1)   \nonumber \\
&\leq& S(A'_1:B'B'_1)  \nonumber \\
&\leq& S(A'_1:B B_1)  = S(A'_1 : B) \nonumber \\
&=& S(A'_1) +S(B) -S(A'_1 B)   \\
&=& H\{ p_i ;i\} + S(\rho_B) -S(\sum_i p_i \ket{i} \bra{i} \otimes \rho_i^B) \nonumber \\
&=& S(\rho_B) -\sum_i p_i s(\rho_i^B) \nonumber \\
&\leq & S(\rho_B) \nonumber \edeqn where $ p_i \rho_i^B = Tr_{A} (
K_i^A \rho_{AB} K_i^{A\dagger} )$, and $\rho_B= \sum_i p_i \rho_i^B
$. Here the first inequality follows from the fact that the quantum
mutual information does not increase by discarding a subsystem, and
the second inequality follows from the fact that quantum mutual
information does not increase under local operations.  Similarly we
can get the symmetric relation: \bgeq I\{A:B\} \leq S(\rho_A) .  \edeq
On the other hand we also have \bgeqn
I\{A:B\} &=   & S(A'_1:B'_1)   \nonumber \\
&\leq& S(A'A'_1:B'B'_1)  \nonumber \\
&\leq& S(A A_1:B B_1)  \nonumber \\
&=& S(A:B) . \nonumber \edeqn Therefore the proposition is proved.


\end{document}